# Towards Understanding Enablers of Digital Transformation in Small and Medium-Sized Enterprises

## Research-in-progress

**Sachithra Lokuge**
School of Business
University of Southern Queensland
Queensland, Australia
Email: ksplokuge@gmail.com

**Sophia Xiaoxia Duan**
School of Accounting, Information Systems and Supply Chain
RMIT University
Victoria, Australia
Email: sophia.duan@outlook.com

## Abstract

Even though, digital transformation has attracted much attention of both academics and practitioners, a very limited number of studies have investigated the digital transformation process in small and medium-sized enterprises (SMEs) and the findings remain fragmented. Given the accessibility and availability of digital technologies to launch digital transformation initiatives and the importance of SMEs in the economy, a profound understanding of enablers of the digital transformation process in SMEs is much needed. As such, to address this, in this paper we conducted a comprehensive review of related literature in information systems, management, and business disciplines, to identify key enablers that facilitate the digital transformation process in SMEs.

# 1  Introduction

The advent of digital technologies such as social media, mobile technologies, analytics, cloud computing, and internet-of-things has given rise to digital transformation (Lokuge et al. 2019; Vial 2019). As per Warner and Wäger (2019), digital transformation is the use of digital technologies to create value-added products and processes in firms and integrate them into their processes, structures, and working models. In recent times, digital transformation has attracted much attention among both academics and practitioners as firms are embracing such initiatives for achieving sustainable competitive advantages in response to the intense digital disruption (Lokuge et al. 2021; Vial 2019). While prior research on digital transformation has focused on large organisations, studies investigating digital transformation in small and medium-sized enterprises (SMEs), remains limited and fragmented (Garzoni et al. 2020; Li et al. 2018). A profound understanding of the digital transformation process in SMEs will help to develop feasible strategies and policies for SMEs to embrace the enormous benefits that digital technologies have provided. Such initiatives will in turn contribute to the sustainable growth of the economy both nationally and globally. As such, it is deemed necessary to understand the enablers of the digital transformation process for SMEs.

SMEs are a distinct group of organisations with unique characteristics that are often reflected from characteristics such as lack of technical expertise, poor infrastructure, inadequate capital, inadequate organisational planning, strong influence of the owner on decision making, extreme dependence on business partners, limited resources, and the presence of greater external uncertainty (Duan et al. 2012; King et al. 2014; Deng et al. 2019). On the other hand, SMEs are more responsive, flexible, and agile in response to digital disruption (Ghobadian and Gallear 1997; Li et al. 2018). As such, organisational theories, and practices applicable to a large firm may not be suitable or applicable for an SME context (Szopa and Cyplik 2020). Such characteristics warrant the need for a separate investigation of the enablers of digital transformation process in SMEs. As such, with the increasing popularity of digital transformation and unique features of SMEs in response to the digital disruption, a review of the related literature could fill the gap and inform research in this domain. To investigate this phenomenon, the overarching research question developed in this study is: *What are the enablers of the digital transformation process in SMEs?*

To address the research question, a systematic review of the related literature in information systems, management, and business disciplines on digital transformation in SMEs was conducted. The multidisciplinary nature of the topic digital transformation deemed such investigation. A total of 46 articles from information systems, management, and business disciplines were synthesized to understand the enablers of digital transformations in SMEs. The findings of the literature review will provide us with a sense of current research and trends on digital transformation in SMEs and this will further set the foundation for identifying future research areas.

The structure of the paper is as follows. The next section describes the research methodology followed in conducting the literature review. Then, in the next section, the findings and the research framework derived through the analysis are provided. Finally, we conclude our paper by highlighting a future research agenda, limitation and concluding remarks.

# 2  Methodology

To investigate the enablers of digital transformation of SMEs an extensive literature review was conducted. The advantages of such an approach are (i) it enables us to get an understanding of the status of the current literature, and (ii) it helps us to identify research gaps. When conducting the review, we did not specify a period. We included all literature that described digitalization, digital transformation in SMEs. For the literature review, we selected journal papers from databases such as AIS Library, INFORM, ProQuest, Wiley, Emerald and EBSCOhost. The selection of the papers preferred inclusion, over exclusivity. The following keywords were used for identifying papers, "digital transformation," OR "digitalization," OR "digitization," OR "digital disruption" AND "Small business," "SME," "Small and medium enterprise." This initial search identified 93 papers that included the above terms. The two researchers went through the sample papers and by reading the title, keywords, and a general search, removed 38 articles. From the remaining 55 articles, the abstract of each of the articles were read and duplicate publications and irrelevant publications were removed. From this, 7 publications were removed as they were not relevant to the topic of analysis. Another 2 were removed due to duplication. Then, the remaining articles (46 articles) were read in their entirety and the relevance was determined. Table 1 provides details of the sample. Due to the page limitation, all references are not provided.

The authors independently analysed each of the papers and identified the factors that determined the success of the digital transformation process in SMEs. Each author identified key themes that they thought were emerging through the analysis. This approach provided a continuous free-flowing mental state in which to absorb the phenomenon of interest. Each paper was analysed separately to identify key enablers and labelled any important information in the process until the existing labels were repeated. Then, the researchers discussed, explored, and refined the relationships between the enablers that were identified independently.

| Year | References | # of papers |
|------|-----------|-------------|
| 2021 | (Beckmann et al. 2021; Del Giudice et al. 2021; Dutta et al. 2021; Garzella et al. 2021; Rivera-Trigueros and Olvera-Lobo 2021) | 11 |
| 2020 | (Balakrishnan and Das 2020; Crupi et al. 2020; Depaoli et al. 2020; Garzoni et al. 2020; Wewege et al. 2020) | 13 |
| 2019 | (Bouwman et al. 2019; Chan et al. 2019; Garbellano and Da Veiga 2019; Llinas and Abad 2019; Nair et al. 2019; North et al. 2019; Pelletier and Cloutier 2019; Riera and Iijima 2019; Sehlin et al. 2019) | 11 |
| 2018 | (Chester Goduscheit and Faullant 2018; Sousa and Wilks 2018) | 4 |
| 2017 | (Scuotto et al. 2017) | 1 |
| 2016 | (Ansari et al. 2016; Ojala 2016) | 2 |
| 2015 | (Prindible and Petrick 2015; Taiminen and Karjaluoto 2015) | 2 |
| 2014 | (Mehra et al. 2014) | 1 |
| 2013 | (Mathrani et al. 2013) | 1 |

*Table 1. Details of the Literature Review Sample*

## 3   Analysis, Findings and Discussion

There is an increased interest in studying digital transformation in SMEs among scholars, exemplified by the increased number of publications each year from 2013 ($n$=1) to 2021 ($n$=11). Even though a period was not specified for the search, we found the papers on digital transformation in SMEs written from 2013. A few papers ($n$=18) focused on a particular industry such as banking, construction, manufacturing, IT, and oil and gas. In the sample, it was identified that most studies ($n$=17) adopted a qualitative approach, followed by the quantitative approach ($n$=7), mixed method ($n$=8), and literature reviews ($n$=3). Maturity model was the most used theoretical framework in the sample. In addition, dynamic capabilities, resource-based view, absorptive capacity, and technology, organisation, and environment framework were commonly used theories and frameworks in understanding digital transformation in SMEs.

### 3.1   Towards Understanding Enablers of Digital Transformation in SMEs

The accessibility of digital technologies and the recent COVID-19 situation has accelerated the digital transformation initiatives among organisations (Argüelles et al. 2021; Lokuge et al. 2020; Vial 2019). While Vial (2019, p. 118) posits digital transformation as "a process that aims to improve an entity by triggering significant changes to its properties through combinations of information, computing, communication, and connectivity technologies," Wessel et al. (2020), argues that digital transformation process triggers the emergence of new organisational identity. As such, the resources, methods, strategies that organisations used to follow, radically need to be changed to enable the digital transformation processes (Wessel et al. 2020). The analysis of literature on digital transformation in SMEs highlighted several important aspects of the process. For example, it highlighted the influence of external environment on digital transformation process of SMEs. The pressure that was exerted from the technological advancement, market demands, lifestyle changes and unprecedented events (i.e., pandemic) mandated digital transformation initiatives in the organisation (Lokuge and Sedera 2014b; Lokuge and Sedera 2014c). The initial analysis of the sample identified 48 enablers. The two researchers independently categorized the factors by merging similar factors and removed any duplication of codes. This resulted in identifying six broad factors that influence the digital transformation process in SMEs.

**Organisational Strategy that favours Digital Transformation:** Organisational strategy is defined through specified goals, objectives, policies and plans for a particular organisation (Miles et al. 1978). Nair et al. (2019) highlighted the importance of clarity of goals for an SME to initiate digital transformation projects. As per Pelletier and Cloutier (2019) when determining a digital strategy, it is important to identify areas where the organisations need to position their resources, identify their capabilities and leverage such capabilities. Due to resource limitations that SMEs face, it is important for them to assess such capabilities for developing a strategy that aligns with their business processes and IT resources (Lokuge and Sedera 2018; Nwaiwu et al. 2020). In addition, the importance of leadership intervention in managing and aligning such organisational strategies was also deemed important (Lokuge et al. 2018b; Philipp 2020; Szopa and Cyplik 2020; Chong and Duan 2020).

**Sustainable Technology Capabilities:** As per Bharadwaj et al. (2013) IT capabilities can be defined as an organisation's capacity to assemble and utilize its IT-based resources (which includes IT physical resources and IT staff), with other organisational resources and capabilities. A successful digital transformation project will require proper utilization of IT capabilities (Lokuge and Sedera 2020). However, the limitation in resources is one of the main characteristics of SMEs. Mathrani et al. (2013) highlights that for an SME to be successful in digital transformation projects, it is important to manage their IT staff, skills and develop their competencies. Philipp (2020) highlights that considering the limitations of IT resources for SMEs, when they have extensive accessibility to IT resources, that will be a determinant for the successful digital transformation project. However, as per Sedera et al. (2016), when organisations are able to utilize their existing IT resources with complementary IT resources, they are able to minimize such limitations. Such endeavours reiterate the importance of IT knowledge and the capabilities of the staff (Mathrani et al. 2013), which can be limited in the SME context. In addition, with the advancement of new technologies, the expertise required and the technology related knowledge requirements also change (Lokuge and Subasinghage 2020). As such, managing technology capabilities, we identify as 'sustainable technology capabilities' determine the success of digital transformation initiatives.

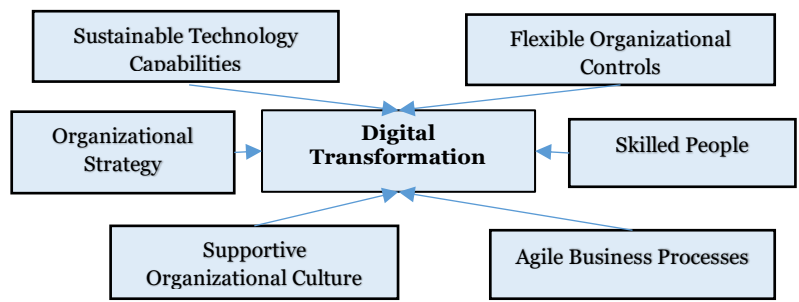

*Figure 1. Key enablers for digital transformation in SMEs*

**Flexible Organisational Controls:** As per Ghobadian and Gallear (1997), SMEs have flexible organisational hierarchy, high top management visibility, close top management intervention, low degree of formalization and high personal authority. Such characteristics as per Jansen et al. (2006) epitome enablers of innovation. As such, theoretically, digital transformation projects can be successfully initiated in SMEs due to their flexibility in organisational structures and controls. However, Bodewes (2002) highlights that formalization is a means of coordinating and controlling the organisation. As such, if there are no formal coordination and controlling mechanisms, SMEs might encounter negative consequences due to the gravity of digital transformation projects. As such, while the existing organisational structures and controls of SMEs favour innovation (Lokuge and Sedera 2014a; Sedera and Lokuge 2017), managing organisational coordination and control are important for SMEs. Balanced organisational structures and controls, as such, will determine the success of digital transformation initiatives of SMEs.

**Skilled People:** The entity people, includes, knowledge, expertise, leadership, empowerment, and intervention of the managers. Prior literature on innovation, highlights the important role of managers in leading innovation (Duan 2020). Since the digital transformation process is also a radical technological initiative, the role of the managers in managing such initiatives is critical (Kohli and Melville 2019). Similarly, the analysis of the literature sample highlighted the important role of managers in applying their knowledge, leadership, their managerial practices that assist successful digital transformation projects (Crupi et al. 2020). While top management visibility, close top management intervention and support is high in SMEs, digital transformation initiatives can be managed and launched successfully in the SME context. Li et al. (2018) further highlight that for SMEs to initiate digital transformation projects, the managerial staff needs to renew their innovation cognition and develop managerial social capital to promote such initiatives. As such, it is not only the intervention, manager's knowledge and how they manage such initiatives also determine the success of digital transformation initiatives.

**Agile Business Processes:** Agility of the business processes play a major role in determining the success of digital transformation initiatives. Digital transformation processes introduce new business processes that differentiate and create a new organisational identity (Depaoli et al. 2020; Lokuge et al. 2018a). While such radical approaches will add value to the organisation, it can be argued that the resistance to such radical changes needs to be managed properly, with proper change management processes in place. For SMEs, this is an extremely critical factor as most SMEs do not have standardized, consistent business processes. Such inconsistencies will hinder the digital transformation process. However, flexibility and agility in reinventing such business process will determine the success of digital transformation endeavours.

**Supportive Organisational Culture:** Prior literature highlights that organisational culture is the most important factor that triggers innovation in organisations (Boudreau and Lakhani 2013; Lee et al. 2016; Lokuge et al. 2016). Especially, for digital transformation initiatives that radically change the core of an organisation, managers need to manage this carefully. It is important that SMEs promote pro-innovative culture in the organisation to minimize the resistance to such initiatives. However, considering the SME context, dimensions of culture such as collectivism, uncertainty avoidance, empowerment, power distance associate positively with innovation (Çakar and Ertürk 2010). Managing such aspects will determine the success of digital transformation initiatives in SMEs.

# 4   Conclusion and Future Work

The objective of this paper was to understand the enablers of the digital transformation process of SMEs. To analyse this phenomenon, a systematic review of 46 papers in information systems, management and business disciplines was conducted. The extensive analysis of the literature sample led to the identification of six key enablers of the digital transformation process in SMEs. While this paper opens new pathways for IS researchers, empirical research is required to establish the initial findings of this study. Our future work will focus on testing and validating these enablers to provide SMEs with clear, empirical evidence and managerial implications for designing effective digital strategies for successful digital transformation projects. The current findings will be important for managers of SMEs to understand and determine the factors for successful digital transformation initiatives. While we acknowledge the size of the literature sample, we only considered journals for our analysis. Conference publications were not considered for the analysis as we focused on established empirical and theoretical research work. However, as an extension to this literature review, we will be including major IS, and management conference papers as well.

## Copyright